\magnification=\magstep1
\settabs 18 \columns

\hsize=16truecm
\baselineskip=17 pt

\def\b{\bigskip}
\def\bb{\bigskip\bigskip}

\def\no{\noindent}
\def\r{\rightline}
\def\ce{\centerline}
\def\ve{\vfill\eject}

\def\r{\rightline}
\font\got=eufm8 scaled\magstep1

\def\~{\hskip-1mm}
\def\g{\hbox{\got g}}
\def\h   {\hbox{\got h}}
\def\k{\hbox{\got k}}

\def\harr#1#2{\smash{\mathop{\hbox to .25 in{\rightarrowfill}}
 \limits^{\scriptstyle#1}_{\scriptstyle#2}}}

\font\aab=cmbsy10 at 18pt
\def\bigbullet{\lower.3ex\hbox{{\aab  \char'017}}}

\def\today{\ifcase\month\or January\or February\or March\or
April\or  May\or June\or July\or August\or September\or
October\or November\or  December\fi \space\number\day,
\number\year }

\r { CERN-TH/2000-152}
\r { UCLA/00/TEP/18}

\r \today
\b





\ce{{\bf CONFORMAL FIELDS IN HIGHER DIMENSIONS}
\footnote*{To be presented to the International Conference
dedicated to the memory of Professor Efim Fradkin, Moscow June
5-10, 2000,  and to the Ninth Marcel Grossmann meeting, Rome July
2-8 2000.}}
\vskip1cm
\ce{S. Ferrara$^{\dagger\ddagger}$ and C. Fr\o nsdal$^\dagger$}
\vskip1cm
\ce {$^\dagger$ Physics Department, University of California Los
Angeles CA 90095-1547, USA }
\vskip.5cm
\ce {$^\ddagger$ CERN Theoretical Division, CH 1211 Geneva 23
Switzerland}
\vskip1cm
\no {\it ABSTRACT.} We generalize, to any space-time dimension,
the unitarity bounds of highest weight UIR's of the conformal
groups
with Lie algebras   $so(2,d )$.  We classify gauge theories
invariant under $so(2,d)$, both integral and half-integral spins.
A similar analysis is carried out for the algebras $so^*(2n)$.

We study new unitary modules of the conformal algebra in $d>4$,
that have no analogue for $d\leq 4$ as they cannot be obtained
by  ``squaring" singletons. This may suggest the interpretation of
higher dimensional non-trivial conformal field theories as
theories of ``tensionless" $p$-branes of which tensionless strings
in $d=6$ are just particular examples.
\vskip1cm

\line{{\bf Introduction.} \hfil}
\vskip.3cm

Extensive work on $AdS_{d+1}$ and its relation to conformal field
theories on ${\cal{M}}^c_d = \partial~AdS_{d+1}$
has found
an interesting realization in supergravity and string
and $M$ theory by relating the horizon geometry of $p$-branes to
the world-volume dynamics of the brane  [GiT].
The conjectured duality [Mal][GKP][W1] between theories of gravity
and boundary conformal field theories is particularly
powerful in the case of space-time supersymmetric field theories
where the
$p$-brane can be ``BPS" saturated; that is, when its world volume
dynamics preserves some  fraction $ \leq N/2$ of the original $N$
supersymmetries.  Theories with maximal supersymmetry correspond
to
$p= d-1 =2,5$ and 3 respectively in
$M$ theory and in IIB string theory compactified on
$AdS_{d+1}\times S_{D-d-1}$ [GuT].  In these theories a peculiar
phenomenon already occurs for $D=11,~d=6$ where the world-volume
$M$-theory five brane (2,0) six-dimensional field theory is
believed to be a non-trivial and interacting theory of
``tensionless self-dual strings" [W2][SW][Se][St]. The $AdS_7/
CFT_6$ correspondence [Mal] predicts  that such a theory, at least
in a certain regime, must be holographically equivalent to 11D
supergravity on
$AdS_7\times S_4$ and in fact certain ${1\over 2}$BPS states of
the latter (such as the K-K towers) [GVNW] can be uniquely
identified with short representations of the (2,0) superconformal
algebra  [AOY][Ha][Mi][ABS] built up by tensoring supersingletons,
 ultrashort UIR's that describe the supermultiplet of
five-brane coordinates transverse to the 5-brane world volume.
[GiT][FeS1,2]

However, the fact that (2,0) conformal field theory is not a theory of
 point particles but is believed instead to be a theory of
``tensionless strings" [W2][SW] should be reflected in the spectrum of
``observable" conformal fields, possibly not the same as those
that are detected in supergravity in $AdS_7$.

It is the aim of the present paper to emphasize a novel feature of
conformal field theories in $d>4$, namely the fact that the spectrum
of ``short" primary conformal fields; that is, the limiting
Harish Chandra modules that become reducible, is wider than what
is naively obtained by ``squaring" singleton representations
[FlFr1] (massless conformal fields) [FFZ]. Tensionless $p$-branes
bring up the subject of combining infinitely many massless fields with
all spins. [Fr7][FV][Gu1][Sz]

A possible interpretation of such new fields is that they are
``currents" related to ``extended objects"; that is, that their
space integrals measures the flux of an extended object in the
boundary conformal field theory.

In the holographic picture such conformal current fields (of higher
rank) should correspond to a new kind of bulk gauge fields in
$AdS_{d+1}$.  Antisymmetric (self-dual when ${d\over 2}$ is
odd) singleton representations in any even dimension were found
by the present authors [FeFr3] together with
a new class of Harish-Chandra
limiting unitary modules for
$d-k$ forms of dimension $E_0 = d-k ~(1\leq k < {d\over 2})$.  All
these modules are ``zero center" modules,  meaning
that all the  Casimirs vanish [FlFr3].

All other singletons (which are not zero center modules) were 
found by Siegel [Si] and also by Minwalla [Mi2] and Angelopoulos and Laoues [AL] by studying general
``massless" conformal fields in $d$ dimensions.  The latter authors also investigated their relations with Poincar\'e and de Sitter groups.  These limiting
Harish Chandra modules correspond to thresholds of the unitarity
bound; the lowest values of $E_0$.

The general problem of classifying all highest weight modules of
the simple Lie algebras was completed by  Enright, Howe and
Wallach [EHW]. Many special cases were known previously.
Here we shall adapt the  results of that paper to the physically
interesting case of the conformal algebras $so(2,d)$.
 We also discuss the algebras $so^*(2n)$. Some of the
symplectic algebras have been studied  already [Fr5][Gu1,2] , while
the unitary and the exceptional Lie algebras do not seem, at this
time, to have found applications in physics. 
In the physics literature, the unitarity bounds of the $so(2,d)$
algebra, corresponding to the limiting Harish Chandra modules classified in [EHW]
were discussed much later: In relation to conformal field theories
they were considered in Ref. [Mi2].  The same algebra 
in connection with gauge fields in $AdS_{2n+1}$,
and their behavior in the Poincar\'e (flat space) limit,
was investigated in [DH] and more recently in [BMV].
References to other special results will be
given.

There is no generally accepted definition of ``masslessness" in
higher  dimension [AFFS][FeFr2][FFG][L][Me2]. We propose that the
most important property to be used for classification of field
theories is whether or not they are gauge theories.   A
universal definition of ``gauge theory" that we think is very
natural is this: A field theory, invariant under a group or a
supergroup
$G$, is a gauge theory if the field or supermultiplet transforms
by a non-decomposable representation of
$G$.  Such representations contain a maximal ideal of states with
zero norm that constitutes the subspace of gauge modes.  In 4
dimensions this property is strongly correlated with masslessness.
(Exceptions: the massless scalar field is not a gauge theory and
singletons are not massless.)  The link between masslessness and
gauge theories is strong in all dimensions whenever
the group $G$ is the conformal group of the manifold.

It will be seen that there are two quite distinct types of ideals
in the limiting Harish Chandra modules. Those that appear as
cases II and III in the enumeration of [EHW] are of the singleton
type.  In odd dimensions ($d$ odd) these are precisely the
two singletons. In the bosonic case, both the Harish Chandra
module and its ideal have highest weights that are trivial on the
semisimple part of the compact subalgebra. In the $AdS_{d+1}$
field theory the full module is carried by the solutions of a
Klein Gordon equation and the physical quotient is distinguished
only by the boundary values. The situation in the fermionic
case is similar. Those listed as types I,p include ordinary
gauge theories of the vector/tensor type (Yang-Mills and
gravity),  in which the ideal appears as exact tensor fields
(gradients). However, most of these are of mixed type. The ideal
is not irreducible and its full characterization  requires some
specification of boundary conditions. The extent of complication
that can arise may be appreciated by examining the case of the
vector singleton in $AdS_5$, as was done in [FeFr1].

\bb

\no{\bf 2. $SO(2,2n)$.   Basis, highest
weight modules.}
\b
\no {\it 2.1. Basis.} In this section  and in the next one
$G_n$, for
$n = 2,3,...$~, is the universal cover of the group
$SO(\delta_n,{\bf R})$, where
$\delta_n$ is a  symmetric, nonsingular 2-form with index (2,2n),
and  $\g_n$ is the associated complexified Lie algebra. The
compact subalgebra $\k_n$ is isomorphic to the direct sum of
  $so(2n,{\bf R})$ and the real,
one-dimensional Lie algebra.

Fix an index set $I =\{0,0',1,...,2n\}$,   $\delta_n$ =
Diag$\{-1,-1, 1,..., 1\}$, and  a basis for $\g_n$,
$$
 \eqalign{&\{L_{ab}= - L_{ba}~;~ a,b\in I, ~a<b \},\cr
&  \hskip-12mm (L_{ab})_c^d = \delta_a^d\delta_{bc} -
a,b,~~~\delta_{bc} := (\delta_n)_{bc},~a,b,c,d \in I.
\cr}
$$
The commutation relations are $[L_{ab},L_{cd}] =
\delta_{bc}L_{da}- a,b - c,d.
$ The compact subalgebra is generated
by
$\{L_{ij}, i,j = 1,...,2n\}$ and $L_{00'}$.

We factor the space $M_{2+2n}$ of $2+2n$ dimensional matrices into
a direct product  $M_2\otimes M_{1+n}$. We introduce the Pauli
matrices
$\sigma_1,\sigma_2,\sigma_3 $ in $M_2$, and the matrices
$$
(e_{ij})_k^l = \delta_i^l\delta_{jk},~i,j,k,l \in \{0,1,...,n\}
$$
 in $M_{1+n}$.

A Cartan
subalgebra $\h_n$ is generated by the set
$$
 h_i = \sigma_2 \otimes e_{ii},~ i = 0,1,...,n.
$$
Positive/negative Serre generators,
$$\eqalign{
& e_{\pm i} = {1\pm\sigma_2\over 2}e_{i,i+1} -
\delta_{ii}{1\mp\sigma_2\over 2}\otimes e_{i+1,i},~ i =
0,1,...,n-1,\cr &
e_{\pm n} = \sigma_1{1\mp\sigma_2\over 2}\otimes e_{n-1,n} -
{1\pm\sigma_2\over 2}\sigma_1\otimes e_{n,n-1},
 \cr }$$
are associated with simple roots $\vec r(j), j = 0,1,...,n$,
linear functions on $\h_n$ defined by
 $
[h_i,e_j] = r_i(j)\,e_j.
$
We find, for $i = 0,1,...,n$,
$$
\eqalign{&
[h_i,e_j] = (\delta_{ij} - \delta_{i,j+1})e_j,~j = 0,1,...,n-1,
\cr & [h_i,e_n] = (\delta_{in} + \delta_{i,n-1})e_n.
\cr}
$$
The positive roots are $r_i(j,k) = \delta_{ij} \pm \delta_{ik},
0\leq j < k \leq n$, and the half-sum of positive roots is
$$
\rho = {1\over 2}\sum \vec r(j,k) = (n,n-1,...,1,0).
$$
Finally, we record the relations
$$\eqalign{
& [e_i,e_{-j}] = \delta_{ij}(h_{i+1}-h_i), ~i,j =
0,1,...,n-1,\cr &
[e_n,e_{-j}] = -\delta_{nj}(h_n + h_{n-1}).\cr}
$$

 \b
\no {\it 2.2. Weights.}  A
`compact weight' will mean a dominant, integral weight on the
Cartan subalgebra
$\h^o_n$ of $so(2n)$ generated by the set $h_1,...,h_n$,
 namely
$$
w_i = w(h_i) = w_i,~ i = 1,...,n,
$$
where $w_1,...,w_n$ are integers or half-integers satisfying
$$
w_1 \geq w_2 \geq ...\geq w_{n-1} \geq
|w_n|.
$$
Each  compact weight $\vec w$ is the highest weight of a finite
dimensional, irreducible representation $D(\vec w)$ of  $so(2n)$.
A `weight' is a pair
$(E,\vec w)$ where $E\in {\bf R}$ is an eigenvalue of $\h_0$ and
$\vec w$ is a compact weight.

\b

\no{\it 2.3. Highest weight modules.} A \g,\k ~module is a
representation of
$\g_n$ on a collection of finite dimensional $\k_n$ modules. A
Harish Chandra module is a
\g,\k~ module that is generated from a highest weight $(E_0,\vec
w)$. Since $E$ will have the interpretation of energy, normally
bounded below, ``highest" will mean that $E_0$ is the lowest
value of $E$. More precisely, consider the decomposition
$$
\g_n = \g_- + \k_n + \g_+,
$$
defined by the grading of $\g_n$  by the adjoint action of
$h_0$. Thus elements of $\g_- (\g_+)$ lower (raise) the energy.
Fix $ (E,\vec w )$ and let $V_0 $ be the associated
 $\k_n$ module, promoted to a
$\k_n + \g_-$ module by letting $\g_-$ act trivially. Then the
Harish Chandra module $V(E,\vec w)$ is the space $U(\g_n)
\otimes'
V_0$, with the natural left action of $\g_n$. The prime on
$\otimes'$ means that $\forall x\in \k_n + \g_-,\, x \,\otimes'  =
\otimes' \,x$.

 Fix a compact weight
$\vec w$. Consider the family of Harish Chandra modules
$V(E_0,\vec w)$ with highest weight
$(E_0,\vec
w), E_0\in {\bf R}$. For $E_0$ large enough this representation
is irreducible and unitarizable. The problem is to determine the
values of $E_0$ such that
\b
(a) The Harish Chandra module is reducible, with a maximal ideal
$I(E_0,\vec w)$, and

(b) The quotient $D(E_0,\vec w) = V(E_0,\vec
w)/I(E_0,\vec w)$ is unitarizable.

\bb
\no{\it 2.4. Results.} Complete results for the case  $n = 2$ were
obtained long ago by physicists. [Mac] The general solution is in
[EHW].
  We need to distinguish a number of different cases.

Let $\Delta_c(\vec w)$ be the set of positive roots $\vec r$
such that $\langle\vec w|\vec r\rangle := \sum_i w_ir_i = 0$.  It
turns out to be one of the following:
\b
Case I,p: $\Delta_c(\vec w)$ is the root system of $su(p), p =
1,...,n$. (When  $p = 1$, then this is the empty set.) Then
$\Delta_c(\vec w)$ contains the roots generated by the simple
roots $\vec r(1),...,\vec r(p-1)$, and this case is characterized
by $\langle
\vec w, \vec r(i)\rangle = 0,\, i = 1,...,p-1$  and $\langle\vec
w,\vec r(p)\rangle \neq 0$, or
$w_1 = w_2 = ... = w_{p} > w_{p+1}$. (Here and below, when $p=n$,
replace the last inequality by $>0$; when $p = n-1$, by $|w_n|$.)

\b
Case II:  $\Delta_c(\vec w)$ is the root system of $so(2n)$.
Then $\vec w = 0$.
\b
The result is that the following is an exhaustive list of
highest weights $(E_0,\vec w)$ that satisfy conditions (a),(b)
above.
\b
Case I,p: $w_1 = w_2 = ...= w_{p} > w_{p+1},\, E_0 = 2n-1 +
w_1-p;~p = 1,...,n$. The lowest energy of the maximal ideal is
$E_0+1$.
\b
Case IIa, IIb: $\vec w = 0, E_0 = n-1$ or 0.The lowest energy of
the maximal  ideal is $E_0 + 2$ (when $E_0 = n-1$) or 1 (in the
other case).
\b
In each case, except the case when $E_0 = 0$, this value  of $E_0$
marks the lower bound for unitary representations, and there are
no other unitary, irreducible representations. The special case
$D(0,\vec 0)$ is the identity representation; it is isolated in
the family of highest weight, unitarizable representations.
\bb

\no {\bf 3. $S0(2,2n)$ invariant gauge theories.}
\b
\no{\it 3.1. Field modules.} Let $M$ be a $G_n$
homogeneous space, ${\cal P}$ a vector bundle over $M$ with
finite dimensional fiber $F$ with a  structure of
$G_n$ module, and $\psi$ a covariant field on $M$, valued in $V$.
We mean by this that we are considering a space $V$ of sections
of ${\cal P}$
that admits an action of $G_n$ induced by the actions on $M$ and
$F$. Without specifying this space in detail, we assume that it is
a $\g,\k$-module, and that there is a subspace that has the
properties of the space $V_0$ in subsection 2.3. Thus $V_0$
is finite dimensional, carries an irreducible representation
of $\k_n$ with highest weight $(E_0,\vec w)$, and is annihilated
by $\g_-$. Note that here $\k_n$ may act only on $M$, only on $F$,
or on both. The problem is to determine whether the action of
$\g_n$ on $V_0$ generates a decomposable $\g_n$ module with a
unitary  quotient. For this to be the case it is necessary, but
not sufficient, that the highest weight be one of the  types
listed in subsection 2.4.\footnote*{ This has been our strategy
 for localizing `` massless" fields in $AdS_4$. See for example
 [Fr1][FeFr1].}

To  proceed it is necessary to choose  the manifold $M$.
The one that is most likely to be of interest is the hyperboloid
$$
\sum \delta_{ab}y^ay^b = 1,
$$
in the pseudo-Euclidean space ${\cal E}_n$ of dimension $2+2n$
with metric
$\delta_n$. Instead of functions on the hyperboloid, it is more
convenient to consider homogenous functions on ${\cal E}_n$. The
vector fields that implement the action of
$\g_n$ are
$$
-L_{ab} = y_a\partial_b - a,b,~ ~y_a := \delta_{ab}y^b,~ a,b \in
I.
$$

It is helpful to introduce a time coordinate $t$, by the polar
decomposition
$$
y_\pm := y^o \pm iy^{o'} =: Ye^{\pm i t}.
$$
Then
$$
h_0 = i{\partial\over \partial t}.
$$
Eigenfunctions of $h_0$ have the form
$(y_+)^\alpha(y_-)^\beta \psi(y_1,...,y_{2n})$ with eigenvalue $E
= \beta - \alpha$.   The subalgebras
$\g_\pm$ (energy raising and lowering operators) are spanned by
$$\eqalign{&
E_j=L_{0j} - iL_{0'j} = 2y_j\partial_+
+y_-\partial_j + \tilde E_j,\cr &
E_{-j}= L_{0j} + iL_{0'j} =  y_+\partial_j + 2y_j\partial_-
+ \tilde E_{-j},\cr}
$$
where $\tilde E_{\pm j}$ are the matrices that represent the
action in $F$.   Choosing the degree of homogeneity equal to
$-E_0$, we obtain a simple representation for the functions
that belong to the ground states. For example, if $F$ is
1-dimensional it is
$$
(y_+)^{-E_0} .
$$

Now we can investigate the Harish Chandra module. From
[EHW] we learn that
the highest weight of the ideal is of the form
$(E_0+1,\vec w')$ in case I and in case II when $E_0 = 0$, and of
the form
$(E_0+2,\vec w')$ in case II when $E_0 = n-1$.
\b
\

\no{\it 3.2. Case I.} The highest weight of the ideal lies in the
space obtained from the ground states by applying one raising
operator. At this level of energy $\k_n$ acts by the
representation $D_{2n} \otimes D(\vec w)$, where $D_{2n}$ is the
defining representation of $so(2n)$. Let
$\{v_r\}$ be a basis for the subspace $V_0$ associated with the
highest weight, orthonomal with respect to the invariant inner
product and making the matrices of $\k_n$ diagonal. Then
$\{E_i v_r\}$ is a basis for the subspace with energy $E_0+1$.
In this subspace there are vectors of zero invariant norm if and
only if the determinant of the matrix
 $$
M_{ir}^{js} = \langle v_s,E_{-j}\,E_i\,v_r\rangle \propto
\delta_r^s ([E_i,E_{-j}])_{rr}.
$$
vanishes.  (Since
$[E_i, E_{-i}]$ lies in the compact Cartan subalgebra, typically
$[E_i, E_{-i}] = h_{i+1} - h_i$, one easily understands why
repetitions among the components of the  compact weights are
characteristic of the highest weights of reducible Harish Chandra
modules.)
\b
\no{\it 3.3. Case IIa, $E_0 = n-1$.} The highest weight is
$(n-1,\vec 0)$, and the associated function is
$$
f(y) = (y_+)^{1-n}
$$
We apply two raising operators to get
$$
f_{jk} \propto(y_-\partial_j +2y_j\partial_+)y_k(y_+)^{-n}
= \delta_{jk}y_-(y_+)^{-n} - 2n\,  y_jy_k(y_+)^{-n-1}.
$$
The problem is to determine the structure of this space as a
$\k_n$ module. It is a sum of two irreducible representations, one
of them one dimensional and spanned by the trace
$$
\sum_jf_{jj} \propto(\delta_{ab}y^ay^b) (y_+)^{-1-n}.
$$
The first factor is the invariant that is constant on the
hyperboloid and this function is the
highest weight vector of
$D(n+1,\vec 0)$. The subspace of functions generated from this
one is an ideal and is a space of gauge modes, of zero invariant
norm. The Harish Chandra module has the structure
$$
D(n-1, \vec 0) \rightarrow D(n+1,\vec 0).
$$

The best known example is: $n = 2$, the module is a space of
solutions of the covariant Klein Gordon equation in $AdS_5$~
[FeFr1]. If, instead of the hyperboloid, one passes to the cone
(Dirac's cone, conformally compactified Minkowski space), then
the  functions in the ideal vanish and the representation becomes
irreducible. This construction generalizes directly to
AdS$_{2n+1}$~ [FeFr1].
\b

\b
\no{\it 3.4. Case IIb, $E_0 = 0$.} Here $D(E_0,\vec w) = D(0,\vec
0)$ is the trivial representation. The maximal ideal in the
Harish  Chandra module has highest weight $(1, \vec \alpha)$,
where
$\vec\alpha$ is the highest weight of the defining
representation $D_{2n}$ of $so(2n)$ (the vector representation).
There are essentially two different ways that this representation
can appear in a field theory.  (a) The basis for the space of
functions associated with the highest weight is the function 1;
 the trivial representation of $\g_n$ appears as a direct
summand. (b) The basis for the space of functions associated with
the highest weight is the function $t \propto
\log y_+$;  the trivial representation is a proper quotient
of a nondecomposable representation.
Both cases are familiar from the analysis of conformal QED, $n =
2$  [BFH1].

\bb
\no{\bf 4. $SO(2,2n+1)$. Basis, highest
weight modules.}
\b
\no {\it 4.1. Basis.} In this section and in the next one $G_n$,
for
$n = 1,2,...$~, is the universal cover of the group
$SO(\delta_n,{\bf R})$, where
$\delta_n$ is a  symmetric, nonsingular 2-form with index
(2,2n+1), and  $\g_n$ is the associated complexified Lie algebra.
The compact subalgebra $\k_n$ is isomorphic to the direct sum
of   $so(2n+1,{\bf R})$ and the real,
one-dimensional Lie algebra.

Fix an index set $I =\{0,0',1,...,2n,2n+1\}$,   $\delta_n$ =
Diag$\{-1,-1,1,...,1,1\}$, and  a basis for $\g_n$,
$$
 \eqalign{&\{L_{ab}= - L_{ba}~;~ a,b\in I, ~a<b \},\cr
&  \hskip-12mm (L_{ab})_c^d = \delta_a^d\delta_{bc} -
a,b,~~~\delta_{bc} := (\delta_n)_{bc},~a,b,c,d \in I.
\cr}
$$
The commutation relations are $[L_{ab},L_{cd}] =
\delta_{bc}L_{da}- a,b - c,d.
$ The compact subalgebra is generated
by
$\{L_{ij}, i,j = 1,...,2n+1\}$ and $L_{00'}$.
These are square matrices of dimension $2+2n+1$. Let $M_{2+2n+2}$
be the space of matrices obtained by adding a last row and
a last column.  We factor this space
 into a direct product
$M_2\otimes M_{1+n}$. We introduce the Pauli matrices
$\sigma_1,\sigma_2,\sigma_3 $ in $M_2$, and the
matrices
$$
(e_{ij})_k^l = \delta_i^l\delta_{jk}, ~~i,j,k,l \in
\{0,1,...,n,n+1\}
$$
in $M_{1+n+1}$. Finally, we remove the last row and the last
column. The last row of the matrices of dimension $2 + 2n+1$ is
now represented as the collection
$u
\otimes e_{n+1,i}$ and the last column as
$v \otimes e_{i,n+1}$, where $u,v$ are rows, columns of length
2, and  $i$ runs over $\{0,1,...,n\}$.

A Cartan
subalgebra $\h_n$ is generated by the set
$$
 h_i = \sigma_2 \otimes e_{ii},~ i = 0,1,...,n.
$$
Positive/negative Serre generators are
$$\eqalign{
& e_{\pm i} = {1\pm\sigma_2\over 2}e_{i,i+1} -
\delta_{ii}{1\mp\sigma_2\over 2}\otimes e_{i+1,i},~ i =
0,1,...,n-1,\cr &
e_{\pm n} = \big({1\atop \pm i}\bigr) \otimes e_{n,n+1} -
 (1, \pm i)\otimes e_{n+1,n},
 \cr }$$
are associated with simple roots $\vec r(j), j = 0,1,...,n$,
linear functions on $\h_n$ defined by
 $
[h_i,e_j] = r_i(j)\,e_j.
$
We find, for $i = 0,1,...,n$,
$$
\eqalign{&
[h_i,e_j] = (\delta_{ij} - \delta_{i,j+1})e_j,~j = 0,1,...,n-1,
\cr & [h_i,e_n] =  \delta_{in}e_n.
\cr}
$$
The positive roots are $r_i(j,k) = \delta_{ij} \pm \delta_{ik},
0\leq j < k \leq n-1$, and $r_i(j) = \delta_{ij}$, and the
half-sum of positive roots is
$$
\rho = {1\over 2}\sum \vec r(j,k) = (n+{1\over
2},n-{1\over 2},...,{1\over 2}).
$$
Finally, we record the relations
$$\eqalign{
& [e_i,e_{-j}] = \delta_{ij}(h_i-h_{i+1}), ~i =
0,1,...,n-1,\cr &
[e_n,e_{-j}] = -2\delta_{nj} h_n.\cr}
$$

 \b
\no {\it 4.2. Weights.}  A
`compact weight' will mean a dominant, integral weight on the
Cartan subalgebra
$\h^o_n$ of $so(2n+1)$ generated by the set $h_1,...,h_n$,
 namely
$$
w_i = w(h_i) = w_i,~ i = 1,...,n,
$$
where $w_1,...,w_n$ are integers or half-integers satisfying
$$
w_1 \geq w_2 \geq ... \geq
 w_n \geq 0.
$$
Each  compact weight $\vec w$ is the highest weight of a finite
dimensional, irreducible representation $D(\vec w)$ of
$so(2n+1)$.  A `weight' is a pair
$(E,\vec w)$ where $E\in {\bf R}$ and
$\vec w$ is a compact weight.

\b

\no{\it 4.3. Highest weight modules.} See subsection 2.3.
\bb
\no{\it 4.4. Results.} The solution is given in [EHW].  The
special case $n = 1$ was done long ago by Evans  [E]. We need to
distinguish a number of different cases.

Let $\Delta_c(\vec w)$ be the set of positive roots $\vec r$
such that $\langle\vec w|\vec r\rangle := \sum_i w_ir_i = 0$.  It
turns out to be one of the following:
\b
Case I,$p$ ($1\leq p
\leq n$): $\Delta_c(\vec w)$ is the root system of $su(p), p
= 1,...,n$. (When  $p = 1$, then this is the empty set.) Then
$\Delta_c(\vec w)$ contains the roots generated by the simple
roots $\vec r(1),...,\vec r(p-1)$, and this case is characterized by $\langle
\vec w, \vec r(i)\rangle = 0, i = 1,...,p-1$ and $\langle\vec
w,\vec r(p+1)\rangle \neq 1$,   or
$w_1 = w_2 = ... =w_{p} > w_{p+1}$.
\b
Case II:  $\Delta_c(\vec w)$ is the root system of $so(2n+1)$.
Then $\vec w = 0$.
\b
Case III:  $\Delta_c(\vec w)$ is the root system of $su(n),  w_1=
... =
w_n$.
\b
The result is that the following is an exhaustive list of
highest weights $(E_0,\vec w)$ that satisfy conditions (a),(b)
above.
\b
Case I,p: $w_1 = w_2 = ...= w_{p}>w_{p+1},\, E_0 = 2n + w_1-p;~p =
1,...,n$. The lowest energy of the maximal ideal is $E_0 + 1.$
\b
Case IIa, IIb: $\vec w = 0, E_0 = n-{1\over 2}$ or 0. The lowest
energy of the maximal  ideal is $E_0 + 2$ or $1$, respectively.
\b
Case III: $w_1 = ... = w_n = {1\over 2},\, E_0 = n$.
The lowest energy of the maximal ideal is $E_0 + 1$.
\b
In each case, except when $E_0 = 0$, this value  of $E_0$ marks
the lower bound for unitary representations, and there are no
other unitary, irreducible representations.
\bb

\no {\bf 5. $S0(2,2n+1)$ invariant gauge theories.}
\b
\no{\it 5.1. Field modules.} See subsection 3.1.  The problem is
to determine whether the action of
$\g_n$ on $V_0$ generates a decomposable $\g_n$ module with a
unitary  quotient. For this to be the case it is necessary, but
not sufficient, that the highest weight be one of the  types
listed in subsection 4.4.

To  get further it is necessary to choose  the manifold $M$.
The one that is most likely to be of interest is the hyperboloid
$$
\sum \delta_{ab}y^ay^b = 1,
$$
in the pseudo-Euclidean space ${\cal E}_n$ of dimension $2n+3$
with metric
$\delta_n$. See subsection 3.1.

  From
[EHW] we learn that the highest weight of the ideal is of the form
$(E_0+1,\vec w')$ in Case I and in Case III, and in Case II when
$E_0 = 0$, and of the form
$(E_0+2,\vec w')$ in Case II when $E_0 = n-{1\over 2}$.
\b
\no{\it 5.2. Case I.} The highest weight of the ideal lies in the
space obtained from the ground states by applying one raising
operator. At this level of energy $\k_n$ acts by the
representation $D_{2n+1} \otimes D(\vec w)$, where $D_{2n+1}$ is
the defining representation of $so(2n+1)$. See  subsection 3.2.

\b
\no{\it 5.3. Case IIa, $E_0 = n-{1\over 2}$.} The highest weight
is
$(n-{1\over 2},\vec 0)$, and the associated function is
$$
f(y) = (y_+)^{{1\over 2}-n}
$$
We apply two raising operators to get
$$
f_{jk} \propto(y_-\partial_j +2y_j\partial_+)y_k(y_+)^{-{1\over
2}-n} = \delta_{jk}y_-(y_+)^{-{1\over2}-n} - (2n+1)\,
y_jy_k(y_+)^{-{3\over 2} -n}.
$$
The problem is to determine the structure of this space as a
\k ~module. It is a sum of two irreducible representations, one of
them one dimensional and spanned by the trace
$$
\sum_jf_{jj} \propto(\delta_{ab}y^ay^b) (y_+)^{-{3\over 2}-n}.
$$
The first factor is the invariant that is constant on the
hyperboloid and this function is the
highest weight vector of
$D(n+{3\over 2},\vec 0)$. The subspace of functions generated
from this one is an ideal and is a space of gauge modes of zero
invariant norm. The Harish Chandra module has the structure
$$
D(n-{1\over 2}, \vec 0) \rightarrow D(n+{3\over 2},\vec 0).
$$

The best known example is: $n = 1$, the module is a space of
solutions of the covariant Klein Gordon dipole equation in
$AdS_4$, the bosonic singleton [FlFr2], instead of the
hyperboloid, one passes to the cone (Dirac's cone, conformally
compactified Minkowski space), then the  functions in the ideal
vanish and the representation becomes irreducible. This
constrution generalizes directly to AdS$_{2n+2}$.

\b
\no{\it Case IIb, $E_0 = 0$.} Here $D(E_0,\vec w) = D(0,\vec 0)$
is the trivial representation. The maximal ideal in the Harish
Chandra module has highest weight $(1, \vec \alpha)$, where
$\vec\alpha$ is the highest weight of the defining
representation $D_{2n+1}$ of $so(2n+1)$ (the vector
representation). There are   two different ways that
this representation can appear in a field theory, precisely as
described in the case of $so(2,2n)$.   Both cases
are familiar from the analysis of $AdS_4$ QED, $n = 1$ ~ [FH].
\b
\no{\it Case III, $w_1 = ... w_n = {1\over 2}, E_0 = 1$}.
This is other singleton representation. The case $n = 1$ is the
familiar fermionic singleton. In this case the two singletons
combine to a representation of $Osp(1/4)$ [Fr3] and singleton
multiplets combine to form a representation of $Osp(N/4)$ ~ [FN]
[BSST][BD][NST].
For $n>1$, it was discovered by [AL] that there are just two
singletons

\b

\b

\no{\bf 6. $SO^*(2n)$. Basis, highest
weight modules.}
\b
\no {\it 6.1. Basis.} In this section and in the next one $G_n$,
for
$n = 2,3,...$~, is the universal cover of the group
$SO^*(2n)$ of $2n$ dimensional, unimodular matrices that
preserve the hermitean form $\sigma_2\otimes {\bf 1}_n$,  and
$\g_n$ is the associated complexified Lie algebra. The compact
subalgebra
$\k_n$ is isomorphic to $u(n)$.

Fix an index set $I =\{1,...,2n\}$,   $\delta_n$ =
Diag$\{1,..., 1\}$, and  a basis for $\g_n$,
$$
 \eqalign{&\{L_{ab}= - L_{ba}~;~ a,b\in I, ~a<b \},\cr
&  \hskip-12mm (L_{ab})_c^d = \delta_a^d\delta_{bc} -
a,b,~~~\delta_{bc} := (\delta_n)_{bc},~a,b,c,d \in I.
\cr}
$$
The commutation relations are $[L_{ab},L_{cd}] =
\delta_{bc}L_{da}- a,b - c,d.
$ The compact subalgebra is
 the unitary subalgebra.

We factor the space $M_{2n}$ of $2n$ dimensional matrices into
a direct product of $M_2\otimes M_n$. We introduce the Pauli
matrices
$\sigma_1,\sigma_2,\sigma_3 $ in $M_2$, and the matrices
$$
(e_{ij})_k^l = \delta_i^l\delta_{jk}
$$
in $M_n$.

A Cartan
subalgebra $\h_n$ is generated by the set
$$\eqalign{
&h_0 = {1\over 2n}(e_{11} + ... + e_{nn}),\cr &
h_i = \sigma_2 \otimes e_{ii} - 2\h_0,~ i = 1,...,n.\cr}
$$
In this section and in the next one $\delta_{ij}$
is the Kroenecker symbol. Positive/negative Serre generators,
$$\eqalign{
& e_{\pm i} = {1\pm\sigma_2\over 2}e_{i,i+1} -
 {1\mp\sigma_2\over 2}\otimes e_{i+1,i},~ i =
 1,...,n-1,\cr &
e_{\pm n} = \sigma_1{1\mp\sigma_2\over 2}\otimes e_{n-1,n} +
{1\pm\sigma_2\over 2}\sigma_1\otimes e_{n,n-1},
 \cr }$$
(those in the first line compact) are associated with simple roots
$\vec r(j), j = 1,...,n$, linear functions on $\h_n$ defined by
 $
[h_i,e_j] = r_i(j)\,e_j.
$
We find, for $i = 1,...,n$,
$$
\eqalign{&
[h_i,e_j] = (\delta_{ij} - \delta_{i,j+1})e_j,~j = 1,...,n-1,
\cr & [h_i,e_n] = (\delta_{in} + \delta_{i,n-1})e_n.
\cr}
$$
The  positive roots are $r_i(j,k) = \delta_{ij} -
\delta_{ik}, 1\leq j < k \leq n$  (compact)  $r_i(j,k) = \delta_{ij} +
\delta_{ik}, 1\leq j < k \leq n$ (noncompact) and the half-sum
of positive roots is
$$
\rho = {1\over 2}\sum \vec r(j,k) = (n-1,...,1,0).
$$

 \b
\no {\it 6.2. Weights.}  A
`compact weight' will mean a dominant, integral weight on the
Cartan subalgebra
  $su(n)$ generated by the set $h_1,...,h_n$,
 namely
$$
w_i = w(h_i) = w_i,~ i = 1,...,n,~ w_1 + ... + w_n = 0,
$$
where $w_1,...,w_n$ are integers or half integers and $w_1 \geq
... \geq w_n$. Each  compact weight $\vec w$ is the highest
weight of a finite dimensional, irreducible representation
$D(\vec w)$ of
$su(n)$.  A `weight' is a pair
$(E,\vec w)$ where $E\in {\bf R}$ is an eigenvalue of $h_0$ and
$\vec w$ is a compact weight.

\b

\no{\it 6.3.} See subsection 2.3.
\b
\no{\it 6.4. Results.} The solution was found by Enright, Howe and
Wallach  [EHW]. We need to distinguish several cases.
\b

Case I,q, $w_2 = w_3 = ...= w_{q+1}, \,E_0 = 2n -3 + w_2-q;~q =
1,...,n-1$. The lowest energy of the maximal ideal is $E_0 + 1.$
\b
Case II,p: $ w_1 = ... = w_p,~p = 3,...,n,$ with $   E_0 = 2n-3 +
w_1-p,~ p$ even,
$E_0 = 2n-2 + w_1 -p, ~p$ odd.
The lowest energy of the maximal  ideal is $E_0 + [{p\over 2}]$.
\b
In both of these cases, this value  of $E_0$ marks the lower bound
for unitary representations. But here there are additional,
isolated, unitary representations ( for $n  > $), namely

Case 3: Same as case II, except that $E_0$ takes one of the
values
$$
E_0 = 2n + w_1 -2p + 2j,~~0\leq 2j \leq \biggl\{\matrix{p-4,
p~{\rm even}\cr p-5, p~{\rm odd}\cr},
$$
with $j$ integer .  This happens is when $n = p
= 4,~ \vec w = 0,~ E_0 = 0$. The first non trivial case is
$n = 5, p = 4, E_0 = w_1 + 2$.

\b
\no{\bf 7. The six-dimensional case.}

In this section we apply the results to the case of
$AdS_7$, with a 6-dimensional Minkowski boundary, where the
conformal algebra is $so(2,6) \approx so^*(8)$. To label the
highest weight of the compact subalgebra $so(6) \approx su(4)$
we shall use the Dynkin labels $a_1,a_2,a_3,$, related to the
highest weight $\vec w$ in the following way:
$$
2w_1 = a_1 + a_3 + 2a_2,~ 2w_2 = a_1 + a_3,~ 2w_3 = a_3 - a_1.
$$
The highest weight of the Harish Chandra modules will be
indicated by the quadruple $E_0, a_1. a_2, a_3$. The three
first classes of unitary representations listed in subsection 2.4
as Cases $I,p  = 1,2,3$ are:
$$\eqalign{
&p = 1: E_0 \geq 4 + w_1 = 4 + a_2 + (a_1+a_3)/2,~~a_2 \neq 0,\cr
& p = 2: E_0 \geq 3 + w_1 = 3 + (a_1+a_3)/2,~~a_2 = 0,\cr &
p=3: E_0 \geq 2 + w_1 = 2 + a_1/2,~~ a_2 = a_3 = 0,~ {\rm or}
~1,3 \rightarrow   3,1.\cr}
$$
The irreducible representations at the bound of the third class
are the singletons, found in [Si][Mi2][AL] and associated with massless
fields on the boundary. Later they were discussed, in the context
of the $Osp(8^*/4)$ superalgebra, in the papers [GuT],[FeS1].
In a recent paper the extension of the unitary
modules corresponding to $p=1$ and $p = 3$, to the superalgebra
$Osp(8^*/4)$, was found  [FeS2].

Notice that the bound in the first case is twice the bound in
the third case. The limiting Harish Chandra modules in Case $p=1$
can be constructed by squaring the singleton representations.
This is the usual situation, where massless fields in the bulk
correspond to (conserved) tensor currents on the boundary,
that are bilinears in boundary massless fields  [FeFr2]. These two
series,
$p = 1,3$ (for $a_3 = 0$)  were investigated in [FeFr3]
and in [FeS1].

The series in the intermediary case, $p = 2$, has no analogue in
4 dimensions. What is new here, in six dimensions, is that, for
$p = 2$, the unitary bound  on $E_0$ is lower than the conformal
degree of the conserved singleton currents. The fields associated
with the unitary bound in case $p=1$ are neither
elementary massless, nor composite. The singletons with $E_0 =
3,\, a_1 = 2$ ($p=3$), and the operators with $E_0 = 4,\, a_1 = a_3
= 1$ ($p = 2)$ were discussed in [FeFr3].

The unitary quotient of the Harish Chandra module $D(E_0, 1,0,1)$,
with $E_0 = 4,\break  a_1= a_3 = 1, a_2 = 0$ can be represented as a
closed 4-form, $dJ_4 = 0$, or its conserved dual $J_4^* = J_2$,
$$
(d^*J_2)_\mu = \partial^\nu J_{\mu\nu} = 0.
$$
The integral
$$
\int_{{\cal M}_4} J_4 = Q_1
$$
defines the flux of a string in six dimensions, so $J_4$ is the
current operator that is related to ``tensionless"
strings [W2][SW][Sa][DFKR][DLP].

This also explains why\footnote*{Moreover, the string is self
dual, which means that it is not only dyonic, but that its
electric and magnetic charges are equal. In this respect there is
a distinction between even and odd values of $n$, since for odd
$n$ the singleton representation $p = n$ is real, while for
even $n$ it is complex.}
we do not have a simple candidate for such an operator; it exists
as a consequence of the tensionless string interaction, and it
cannot be described as a local, bilinear in the massless
(singleton) fields.

The picture presented here suggests that there could be
a bulk theory that generalizes supergravity by including this new
bulk field, with interactions.\footnote*{However, the unitarity bounds
of $Osp(8^*/2N)$ superalgebras seem to exclude this possibility [Mi2][FeS2].}

Such a current $J_4$ is actually known in 6-dimensional (1,0)
supersymmetric theories [Sa] where, away from the conformal point,
tensor multiplets interact with non abelian gauge fields and
it takes the form [Sa][FRS][DFKR]
$$
J_4 = {\rm Tr} \,F\wedge F, ~~ Q_1 = \int{\rm Tr} F\wedge F.
$$
The string flux is related to the instanton number (instanton in
a space transverse to the string). [DLP] The flat limit of a
tensionless string was discussed in [DLLP].
 \bb

\no{\bf 8. Speculations about higher dimensions.}

For higher dimensions, with $d-2$ a multiple of 4, we may think
that the above formulas generalize to ($d = 2n$)
$
J_{n+1} = {\rm Tr} F^{{n+1\over 2}},
$
so that
$$
Q_{n-2} = \int_{{\cal M}_{n+1}} {\rm Tr} F^{{n+1\over 2}},
$$
which is the ${n+1\over 2}$ Chern class of the gauge group.

The unitary bound for the family $D(E_0, \vec w)$ of Harish
Chandra $so(2,2n)$ modules, for a fixed, integral, dominant
compact weight $\vec w$ was obtained in subsection 2.4.
We concentrate on the cases,
$$
w_1 = ... w_p > \matrix {w_{p+1},~ p<n,\cr      0, ~~~  p = n\cr}
,
$$
where the bound is
$$
E_0 = 2n-1 + w_1 -p.
$$
Singletons are in the class $p = n$.

The particular case $w_1 = 1$, $E_0 = d-p$, was investigated in
[FeFr3]. The fields are closed $d-p$ forms,
$$
d J_{d-p} = 0.
$$
Indeed, this equation is conformally invariant [W1] if the
conformal degree is $d-p$.

The integral
$$
Q_{p-1} = \int_{{\cal M}_{d-p}} J_{d-p}
$$
is the flux of a $p-1$ brane ($d-p$ are the coordinates transverse to
the brane) that we
interpret as tensionless $p-1$-brane, a natural generalization
from the 6-dimensional case.

Let us consider a hypothetical $d= 10$ dimensional conformal
field theory, the holographic description of a hypothetical 11
dimensional theory in $AdS_{11}$. [Gu1][Ho] The bosonic singleton,
other than the scalar, is a self dual five form with $E_0 = 5$.
However, there is a whole set of unitary Harish Chandra
modules that are all zero center modules (all the Casimir
operators of the conformal group vanish on them), described
by a 4-form with $E_0 = 6$, a 3-form with $E_0 = 7$, a 2-form
with $E_0 = 8$ and a vector with $E_0 = 9$. ($ p = 1,2,3,4
$.)
The integrals of these currents may produce fluxes for 3,2,1 and
0 branes, respectively. The last is the usual global charge
present in all theories (for any $d \geq 3$). Therefore, a 10
dimensional conformal field theory may be considered as a theory
of tensionless 3,2, and 1 branes [FP].

It is tempting to go on to suggest that the only conformal branes
are the dyonic (self dual) 3-branes, analogues of the dyonic
string in dimension 6. In support of this there is the obvious
fact, already observed in [FeFr3],
that this theory resembles a kind of conformal limiting case of IIB
supergravity.  In analogy with the 6 dimensional case
we may think that such a theory, away from the conformal point,
is defined thropugh non abelian gauge interactions, so that the
6-form  current is of the type
$$
J_6 =  {\rm Tr} F^3,
$$
where $F$ is a non abelian 10 dimensional gauge field. In such a
situation the 3-brane charge would be related to a topological
configuration of the gauge field with non vanishing third Chern
class
$$
Q_3 = \int{\rm Tr} F^3.
$$

It is not known whether the above considerations may be accommodated in a
supersymmetric theory.  To answer this question further studies on
new types of supersymmetric structures in higher dimensions may be needed.

\ve
\no{\bf Appendix. Case by case.}
\b
\no{\it A.1. SO(2,2n).}

\underbar {$ n=1.$} Since $SO(2,2)$ is not simple, this case is
not encompassed by the investigation of [EHW]. The results apply
nevertheless if we interpret as follows. Take a standard
Chevalley basis for each of the two factors $SO(2,1)$, with
generators $A,B$ of the compact Cartan generators normalized so
that the weights are $\pm 1$ in the adjoint representation.
Set  $h_0 = (A+B)/2,\, h_1 = (A-B)/2$. Let $(a,b)$ be a generic
pair of eigenvalues of $(A,B)$, then our highest weight is given
by $E_0 = (a+2)/2,\, w_1 = (a-b)/2$.
\b
Case I,1 is the case $b=1, E_0 = w_1 > 0$. The Harish Chandra
module is irreducible on the first $SO(2,1)$ factor and
equivalent to $D(0) \rightarrow D(1)$ on the other. This
representation appears in the gauge theory of singletons in
$2+1$ dimensions  [Fr6][FlFr4].
\b
Case II is  $ a = b = 0, \,E_0 = w_1 = 0$. The quotient is
$D(0,0)$ and the ideal is $D(1,1), D(1,-1)$ or both.

\b
\underbar{$n = 2$.}~ $SO(2,4)$ is the anti De Sitter group in 5
dimensions and the conformal group in 4 dimensions. The compact
subgroup is $SO(4) = SU(2) \otimes SU(2)$. The usual notation
for the compact weights is $(J_1,J_2)$ and is adapted to this
decomposition, with $J_1,J_2$ integral or half integral. Set
$w_1 = J_1 + J_2,\, w_2 = J_1 - J_2$.
\b
Case I,1 is the general case, $w_1 > w_2$ or $J_1J_2 \neq 0$. The
formula
$E_0 = 2n-1+w_1-p$ becomes $E_0 = J_1 + J_2 + 2$. The lowest
energy of the ideal is $E_0+1$.
The simplest field theory in $AdS_5$, with $J_1 = J_2 = {1\over
2}$, is the theory of a vector field, homogeneous of degree
$-3$. The ground states are
$$
f_i(y) = (y_+)^{-3}\epsilon_i,~~i = 1,2,3,4,
$$
with $\epsilon_i$ constant. The lowest  weight of the ideal is
$(4,0,0)$ and the associated gauge field subspace consists of the
gradients or exact vector fields. The next simplest cases
are the  fields of $AdS_5$ supergravity, with highest weights
$(E_0,w_1,w_2)$ equal to $(4,1,1)$ and $({7\over 2}, {1\over
2},1), ({7\over 2}, 1,{1\over 2})$.
\b
Case I,2 is characterized by $w_1 = w_2$ or $J_2 = 0$. The
formula for $E_0$ becomes $E_0 = J_1 + 1$. This is the familiar
case of conformally invariant field theories in 4-dimensional
Minkoski space (or $AdS_4$)  [BFH1].  Realized as field theories in
$AdS_5$ they are topological singleton field theories  [FeFr1].
\b
Case IIa has $w_1 = w_2 = 0$ and $ E_0 = 1$. It is the
representation associated with a scalar, massless field in 4
dimensions. Case IIb is the case
$E_0 = J_1 = J_2 = 0$. The ideal has highest weight $(1,1,0)$; it
is the highest weight of a non unitary irreducible representation.

\b
\underbar {$n = 3$.} ~$SO(2,6)$ is the  symmetry group of $AdS_7$.
See Section 7.

\bb
\no{\it A.2. SO(2,2n+1).}

\underbar {$n = 1.$} The nondecomposable representations of
$S0(2,3)$, and the associated field theories in $AdS_4$ have been
studied extensively.
\b
Case I,1: $w_1 > 0,\, E_0 = w_1 +1$. The highest weight of the
ideal is $w_1+2,w_1-1$. These are the representations
associated with massless fields with spin greater than or equal
to 1   [BFH2,3][[FFr1][Fr2][FH].
\b
Case IIa:~ $w_1 = 0,\, E_0 = {1\over 2}$. The highest weight of
the ideal is $({5\over 2},0)$. This is the bosonic singleton,
described by a scalar field  [FlFr2].
\b
Case IIb: The Harish Chandra module is $D(0,0) \rightarrow
D(1,1)$; the ideal non unitary. This is a component of the field
representation of QED in $AdS_4$. [BFH2]
\b
Case III: $w_1 = {1\over 2}, \, E_0 = 1$; the fermionic
singleton, described by a spinor field  [He][P].

\b

\underbar {$n = 2.$} $SO(2,5)$ is the symmetry group of $AdS_6$.
\b
Case I,1: $w_1 > w_2,\, E_0 = 3 + w_1$.

Case I,2: $w_1 = w_2 >0,\, E_0 = 2 + w_1$.

Case IIa,IIb: $\vec w = 0,\, E_0 = {3\over 2}$, the bosonic singleton,
or $0$.

Case III: $w_1 = w_2 = {1\over 2},\, E_0 = 2$, the fermionic singleton.
This case is
particularly interesting since it is related to the exceptional F(4)
superconformal algebra [N][R]. The spin zero and spin one-half
singletons are combined in the singleton hypermultiplet.

\bb

\no{\it A.3. SO$^*$(2n).}

The isomorphisms $so^*(4) = so(2,1) \times so(3),~ so^*(6) = su(1,3)$ and
$so^*(8) = so(2,6)$ allow us to omit the cases $n = 2,3,4$.  We do not 
discuss the $n>4$ cases.

\bb
 \no{\bf Acknowledgements.}

The work of S.F. has been supported in part by the European
Commision TMR programme ERBFMRX-CT96-0045 (Laboratori Nazionali
di Frascati, INFN) and by DOE grant DE-FG03-91ER40662, Task C.

\bb

\no{\bf References.}

 \item {[ABS]} Aharony, O., Berkooz, N. and Seiberg, N.,
``Lightcone description of (2,0) superconformal theories in six
dimensions", Adv. Theor. Math. Phys. {\bf 2} (1998) 119.
(hep-th/9712117)

 \item {[A0Y]} Aharony, O., Oz, Y. and Yin, Z., ``M-theory on
$AdS(p) \times S(11-p)$ and superconformal field theories", Phys.
Lett. {\bf B430} (1998) 87. (hep-th/9803051)

 \item {[AL]} Angelopoulos, E. and Laoues, M., ``Massless in $n$
dimensions", Rev. Math. Phys. {\bf 10} (1998) 271.
(hep-th/9806199)

 \item {[AFFS]} Angelopoulos, E., Flato, M., Fronsdal, C. and
Sternheimer, D., `` Massless particles, conformal group and De
Sitter Universe", Phys. Rev., {\bf D23} (1981) 1278.

\item {[BSST]} Bergshoeff, E.D., Salam, A. Sezgin, E. and Tanii,
Y. `` Singletons, higher spin massless states and the super
membrane", Phys. Lett{\bf B205} (1998) 137.

\item {[BD]} Blencowe, M.P and Duff, M.J., ``Supersingletons",
Phys. Lett. {\bf B203} (1988) 229.

 \item {[BFH1]} Binegar, B., Fronsdal, C. and  Heidenreich, C.,
``Conformal QED", J. Math. Phys. {\bf 24} (1983) 2828.

 \item {[BFH2]} Binegar, B., Fronsdal, C. and Heidenreich, C., 
``De Sitter QED", Ann. Phys. {\bf 149} (1983)
254.

 \item {[BFH3]} Binegar, B., Fronsdal, C. and Heidenreich, C., 
``Linear, conformal quantum gravity",Phys.
Rev. {\bf D27} (1983) 2249.

 \item {[BMV]} Brink, L., Metsaev, R.R. and Vasiliev, M.A., ``How
massless are massless fields in $AdS_d$,  hep-th/0005136.

 \item {[DFKR]} Duff, M., Ferrara, S., Khuri, R. R. and Rahnfeld, J,
``Supersymmetry in dual string solitons", Phys. Lett. {\bf B356}
(1995) 479.

 \item {[DLLP]} Duff, M. Lu, J.t., L\"u, I.I. and Pope, C.
``Gauge dyonic strings and their global limit", Nuc. Phys. {\bf
B529} (1998) 137.

 \item {[DLP]} Duff, M., Lu, I. and Pope, C., ``Heterotic phase
transitions and singularities of the gauge dyonic string", Phys.
Rev {\bf B378} (1996) 101.

\item {[DN]} Deser, S. and Nepomechie, R. I., ``Gauge invariance
versus masslessness in De Sitter spaces", Ann. Phys. {\bf 154} (1984)
396.

\item {[EHW]} Enright, T., Howe, R. and Wasllach, N., ``A
classification of unitary highest weight modules", in {\it
Representation Theory of Reductive Groups}, P.C. Trombi Ed.\break
Birkh\"auser 1982.

 \item {[E]} Evans, N.T., ``Discrete series for the universal
covering group of the $3+2$ De Sitter group", J. Math. Phys. {\bf
8} (1967) 170.

 \item {[FaFr]} Fang, J. and Fronsdal, C.
``Massless, half-integral spin fields on
De Sitter space", Phy. Rev. {\bf D 22} (1980) 1361.

\item {[FeFr1]} Ferrara, S. and Fronsdal, C., ``Conformal Maxwell
theory as a singleton field theory on $AdS_5$, IIB three-branes
and duality", Class. Quant. Grav. {\bf 15} (1998) 2153-2164.
(hep-th/971223)

 \item {[FeFr2]} Ferrara, S. and Fronsdal, C.,
``Gauge fields as composite boundary
excitations", Phys. Lett. {\bf B 433} (1998) 19. 

 \item {[FeFr3]} Ferrara, S. and Fronsdal, C. 
``Gauge fields and singletons of
$AdS_{2p+1}$", Lett. Math. Phys. {\bf 46} (1998) 157-169.
(hep-th/9806072)

 \item {[FFZ]} Ferrara, S, Fronsdal, C. and Zaffaroni, A., ``
Supergravity on $AdS_5$ and $N = 4$ superconformal Yang-Mills
theory", Nucl. Phys. {\bf B 532} (1998) 153-162.

 \item {[FeP]} Ferrara, S and Porrati, M. ``$AdS$ superalgebras
and brane charges", Phys. Lett. {\bf B458} (1999) 43.

 \item {[FRS]} Ferrara, S., Riccioni, F. and Sagnotti, A,
``Tensor and vector multiplets in 6-dimensional
supergravity", Nucl. Phys. {\bf B 1998} (1998) 115.
(hep-th/9711059)

 \item {[FeS1]} Ferrara, S. and Sokatchev, E.,
``Representations of (1,0) and (2.0)
superconformal algebras in 6 dimensions, massless
and short superfields", hep-th/0001178.

 \item {[FeS2]} Ferrara, S. and Sokatchev, E., 
``Superconformal interpretation
of BPS states in $AdS$ geometry", \break hep-th/0005151.

 \item {[FlFr1]} Flato, M. and Fronsdal, C, ``One massless
particle equals two Dirac singletons", Lett. Math. Phys. {\bf 2}
(1978) 421.

 \item {[FlFr2]} Flato, M. and Fronsdal, C., 
``Quantum field theory of singletons", J.
Math. Phys. {\bf 22} (1981) 1100;
and
``The singleton dipole", Commun. Math.
Phys. {\bf 108} (1987) 469.

 \item {[FlFr3]} Flato, M. and Fronsdal, C., 
`` Spontaneously generated field theories,
zero center modules, colored singletons and the virtues of $N = 6$
supergravity", in {\it Essays
in Supersymmetry}, Reidel, 1986.

 \item {[FlFr4]} Flato, M. and Fronsdal, C., 
``Three-dimensional singletons", Lett.
Math. Phys. {\bf 20} (1990) ?65.

 \item {[FFG]} Flato, M., Fronsdal, C. and Gazeau, J.P.,``
Masslessness and lightcone propagation in De Sitter and 2+1
Minkowski space". Phys. Rev. {\bf D 33} (1986) 415.

 \item {[Fr1]}  Fronsdal, C., ``Elementary particles in curved
space IV. Massless particles," Phys. Rev. {\bf D 12} (1975) 3819.

 \item {[Fr2]} Fronsdal, C., 
``Singletons and massless, integral spin
fields on De Sitter space", Phys. Rev. {\bf D 20} (1979) 848.

 \item {[Fr3]} Fronsdsal, C., 
``The Dirac supermultiplet", Phys. Rev. {\bf
D26} (1982) 1988.

 \item {[Fr5]} Fronsdal, C., 
``Massless particles, orthosymplectic
symmetry and another type of Kaluza-Klein theory",in {\it Essays
in Supersymmetry}, Reidel, 1986.

 \item {[Fr6]} Fronsdal, C., 
``Three-dimensional singletons and
two-dimensional conformal field theory", Int. J. Mod. Phys. {\bf
A7} (1992) 2193, and
``A model for QCD in three dimensions",
in {\it Proceedings of the Colloque Rideau}, Paris 1995.

\item {[Fr7]} Fronsdal, C., 
``Some open problems with higher spins",
Proceedings of the supergravity workshop at Stonybrook, September
1979. P. van Nieuwenhuizen and D.Z. Freedman, Ed.s.

 \item {[FH]} Fronsdal, C and Heidenreich, W., ``Linear De Sitter
gravity", J. Math. Phys. {\bf 28} (1987) 215.

\item {[FN]} Freedman, D. and Nikolai, H., ``Multiplet shortening
in $Osp(N/4)$, Nucl. Phys. {\bf B237} (1984) 342.
 gravitational
interaction of massless high-spin fields", Phys. Lett. {\bf 189B}
(1987) 89./

 \item {[GKP]} Gubser, S.S., Klebanov, I.R. and Poliakov, A.M.,
``Gauge theory correlators from non critical string theory", Phys.
Lett. {\bf B48} (1998) 105. (hep-th/9802109)

 \item {[Gu1]} Gunaydin, M., ``Unitary superalgeras of $Osp(1/32,R)$
and M-theory", Nucl. Phys. {\bf B528} (1998) 432.

 \item {[Gu2]} Gunaydin, M., 
``$AdS.CFT$ dualities and the unitary
representations of non ccompact groups and supergroups: Wigner
versus Dirac", hep-th/0005168.

\item {[GiT]} Gibbons, C.W. and Townsend, P.K., ``Vacuum
interpolation in supergravity via super $p$-branes", Phys. Rev.
{\bf 71} (1993) 3754.

 \item {[GVNW]} Gunaydin, M., van Nieuwenhuizen, P. and Warner, N.,
``General construction of the unitary representations of anti De
Sitter superalgebras and the spectrum of the $S_4$ configuration
of eleven dimensional supergravity", ``Nucl. Phys. {\bf B255}
(1985) 543.

 \item {[GuT]} Gunaydin, M. and Takemae, S. ``Unitary superalgebras of
$Osp(8^*/4)$ and \break $AdS_7/CFT_6$ duality, hep-th/9910110.

\item {[Ha]} Halyo, E., ``Supergravity on $AdS(4/7)  \times S(7/4)$
and M-branes", JHEP 9804,011 (1998). (hep-th/9803077)

 \item {[Ho]}  Horava, P. ``M-theory as a holographic field theory",
Phys. Rev. {\bf D 59} (1989) 046004.

\item {[He]} Heidenreich, W., Nuovo. Cim. {\bf A80} (1984) 220.

 \item  {[L]} Laoues, M., ``Massless particles in arbitrary
dimension",  Math. Phys. {\bf 10} (1998) 1079.

 \item {[Mac]} Mack, G., ``All unitary representations of the
conformal group $SU(2,2)$ with positive energy", Commun. Math.
Phys. {\bf 55} (1977) 1.

 \item {[Mal]} Maldacena, J., `` The large $N$ limit of
superconformal field theories and supergravity", Adv. Theor.
Math. Phys. {\bf 2} (1998) 231. hep-th/9705104.

\item {[Me1]} Metsaev, R. R., ``Arbitrary spin massless bosonic
fields in $d$-dimensional anti De Sitter space", hep-th/9810231.

\item {[Me2]} Metsaev, R. R., 
``Massless mixed symmetry boson and fermion
fields in anti De Sitter space", Phys. Lett. {\bf B354} (1995) 78.

 \item {[Mi]} Minwalla, S., ``Particles on $AdS(4/7) $ and primary
operators of $M(2)$ brane and $M(5)$ brane world volumes", JHEP
{\bf 10} (1998) 002. (hep-th/9803053)

\item{[Mi2]} Minwalla, S., ``Restrictions imposed by Superconformal
invariance on quantum field theories", Adv. Theor. Mat. Phys. {\bf 2}
(1998) 781. (hep-th/9712074)

\item {[N]} Nahm, W., ``Supersymmetries and their representations",
Nucl. Phys. {\bf B 135} (1978) 149.

\item {[NST]} Nicolai, H., Sezgin, E. and Tanii, Y., ``Singleton
representations of $Osp(N/4)$", Nucl. Phys. {\bf B305} (1988) 483.

\item {[P]} Percoco, U., ``The spin ${1\over 2}$ singleton dipole",
Lett. Math. Phys. {\bf 12} (1986) 315.

\item {[R]} Romans, L.J., ``The $F(4)$ gauged supergravity in six
dimensions", Nucl. Phys. {\bf B269} (1986) 691.

 \item {[Sa]} Sagnotti, A., `` A note on the Green-Schwarz
mechanism in open string theory", Phys. Lett. {\bf 294B} (1992)
196.

\item {[Se]} Seiberg, N., ``Nontrivial fixed point of
the renormalization group in 6 dimensions", hep-th/9609061.

\item {[SW]} Seiberg, N. and Witten, E., ``Commment
on string dynamics in 6 dimensions", hep-th/9603003.

\item {[Sz]} Sezgin, E., ``High spin $N = 8$ supergravity in
$AdS_4$, hep-th/9903020.

\item{[Si]} Siegel, W., ``All free conformal representations in all
dimensions", Int. J. Mod. Phys. A {\bf 4} (1989) 2015. 

\item {[St]} Strominger, A., ``Open $p$-branes",
Phys. Lett. {\bf B383} (1996) 44. (hep-th/9512059)

\item {[T]} Townsend, P. K., Nucl. Phys. (Rev. Suppl.) {\bf 68} (91998)
11. (hep-th/9708034)

\item {[W1]} Witten, E., `` Anti De Sitter space and holography", Adv.
Theor. Math. Phys. {\bf 2} (1998) 253. (hep-th/9802150)

 \item {[W2]} Witten, E., 
``Some comments on string dynamics",
(hep-th/9507121)

\end